# Magnetization statics and dynamics in (Ir/Co/Pt)$_6$ multilayers with Dzyaloshinskii-Moriya interaction

A.K. Dhiman[1*], R. Gieniusz[1], P. Gruszecki[2], J. Kisielewski[1], M. Matczak[1], Z. Kurant[1], I. Sveklo[1], U. Guzowska[1], M. Tekielak[1], F. Stobiecki[3], A. Maziewski[1]

[1] Laboratory of magnetism, Faculty of Physics, University of Białystok, Białystok, Poland

[2] Faculty of Physics, Adam Mickiewicz University, Poznań, Poland

[3] Institute of Molecular Physics, Polish Academy of Sciences, Poznań, Poland

*corresponding author: anuj.dhiman@uwb.edu.pl

**Abstract:** Magnetic multilayers of (Ir/Co/Pt)$_6$ with interfacial Dzyaloshinskii-Moriya interaction (IDMI) were deposited by magnetron sputtering with Co thickness $d$=1.8 nm. Exploiting magneto-optical Kerr effect in longitudinal mode microscopy, magnetic force microscopy, and vibrating sample magnetometry, the magnetic field-driven evolution of domain structures and magnetization hysteresis loops have been studied. The existence of weak stripe domains structure was deduced – tens micrometers size domains with in-plane "core" magnetization modulated by hundred of nanometers domains with out-of-plane magnetization. Micromagnetic simulations interpreted such magnetization distribution. Quantitative evaluation of IDMI was carried out using Brillouin light scattering (BLS) spectroscopy as the difference between Stokes and anti-Stokes peak frequencies $\Delta f$. Due to the additive nature of IDMI, the asymmetric combination of Ir and Pt covers led to large values of effective IDMI energy density $D_{eff}$. It was found that Stokes and anti-Stokes frequencies as well as $\Delta f$, measured as a function of in-plane applied magnetic field, show hysteresis. These results are explained under the consideration of the influence of IDMI on the dynamics of the in-plane magnetized "core" with weak stripe domains.

**Introduction:**

A large amount of magnetism research in today's time is concentrated on inhomogeneous magnetic textures, e.g., periodic magnetic textures as stripe domain patterns [1], chiral textures as spin spirals [2], particle-like textures as magnetic skyrmions [3, 4, 5, 6] and various other textures [7]. In addition, in magnonics, particular attention is focused on nonreciprocal media [8, 9], i.e., systems with broken time-reversal symmetry of spin wave (SW) propagation. Particularly interesting in this context are multilayered magnetic systems with perpendicular magnetocrystalline anisotropy and Dzyaloshinskii-Moriya interaction (DMI).

DMI is an anti-symmetric exchange interaction [10, 11] that occurs for materials possessing spatial inversion asymmetry. Initially "bulk DMI" was discovered in B20 weak magnetics due to their structural properties, i.e., broken inversion symmetry [4, 12]. Subsequently, it was found that broken inversion symmetry can be generated artificially for ultrathin ferromagnetic films adjacent to heavy metal materials [13, 14]. Interfaces of ferromagnetic/heavy metal materials produce a symmetry breaking termed as interfacial Dzyaloshinskii-Moriya interaction (IDMI) [15, 16]. It is known that IDMI affects many fundamental static and dynamical magnetic properties - from SW reciprocity to magnetic domain structures [17, 18]. Due to the IDMI, SWs propagating in Damon-Eshbach geometry manifest an asymmetric dispersion relation, and it can be studied using Brillouin Light Scattering (BLS) spectroscopy [17]. Quantification of the value of IDMI strength $D_{eff}$ (effective IDMI constant) and its sign in magnetic multilayers such as Ir/Co/Pt (due to large IDMI) [19] is crucial to understand the behavior of magnetization statics (determination of all the possible magnetic configurations and ways to achieve them) and spin wave dynamics.

In this work, we investigate magnetic multilayers [Ir/Co/Pt]$_6$ where one can expect: (i) strong IDMI and (ii) quality factor (ratio of uniaxial anisotropy to demagnetization energy) $Q<1$. Domain structures were deduced from complementary experimental techniques and supported by micromagnetic simulations. BLS spectroscopy was performed as a function of the magnitude of the in-plane applied magnetic field.

**Methods & Measurements:** Magnetic multilayer of Ir/Co/Pt (nominal structure: Ti(4)/Au(30)/Ir(2)[Ir(1)/Co(1.8)/Pt(1)]$_6$/Pt(2), where thickness in brackets are in nm) was deposited at room temperature using magnetron sputtering on oxidized Si substrate. Magnetization hysteresis loops measurements were performed using vibrating sample magnetometer (VSM) for both in-plane and out-of-plane applied magnetic field. Measurements based on X-band (9.5 GHz) ferromagnetic resonance (FMR) spectroscopy were performed to characterize magnetic anisotropy. The magnetic domain structures were studied at: (i) micrometer resolution using longitudinal magneto optical Kerr effect (LMOKE) microscopy and (ii) tens of nanometers resolution by magnetic force microscopy (MFM) microscopy using low moment CoCr magnetic tips.

BLS spectroscopic measurements were performed in backscattering geometry using green light laser (wavelength $\lambda$ =532 nm). Due to the formation and annihilation of magnons, the frequency positions of Stokes $f_S$ and anti-Stokes $f_{aS}$ peaks were obtained in this inelastic scattering process.

**Results & discussions:** Fig. 1a shows the magnetization hysteresis loops acquired for the out-of-plane (black line) and the in-plane (red line) directions of the magnetic field. Comparing the shapes of these curves one can deduce the in-plane (negative) magnetic anisotropy.

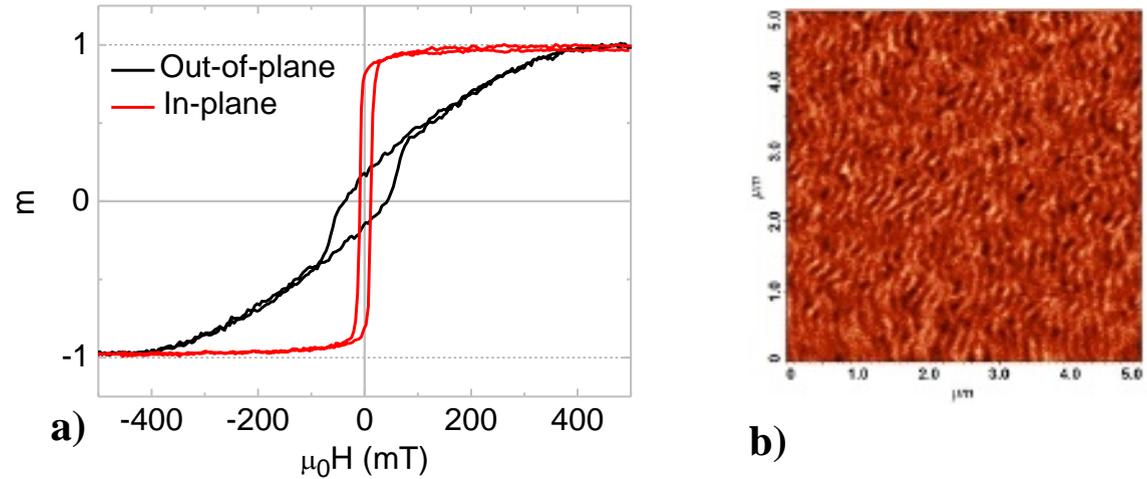

Fig. 1. (a) In-plane (red curve) and out of plane (black curve) magnetization hysteresis loops acquired by VSM; (b) The MFM image of the remanent state ($\mu_0H_y=0$). The image size is 5x5 µm².

It is confirmed after determination of negative uniaxial effective magnetic anisotropy field $\mu_0H_{ueff}$ =-0.45 T, from the X-band FMR spectra measured for magnetic field applied in-plane. Weak stripe domain structure [20, 21] with magnetization distribution modified by IDMI could be expected. Labyrinth pattern of magnetic domains (characterized by out-of-plane magnetization and a period of about 100 nm) is clearly visible in MFM image obtained in the remanent state (Fig. 1(b)).

In-plane field-driven magnetization reversal was studied by LMOKE microscopy, sensitive to in-plane magnetization, see Fig. 2. Initially, the sample was saturated by the field aligned along the +y-axis. Therefore, at the remanence, the in-plane magnetization is aligned along the +y-axis (see the uniformly dark monodomain state in Fig. 2(a)). The following domain evolution was observed while decreasing the value of the magnetic field. Firstly, the nucleation of the domains with the magnetization aligned along the -y-axis and an increase of its area occurs (see the brighter domains emerging and increasing in Fig. 2(b), (c)). Finally, the domains aligned along the +y-axis (the darker domains) vanish, leading to the saturation of the sample along the -y-axis, brighter monodomain state creation. Noteworthy, the magnetic domains (with tens micrometer size) of irregular shape are elongated along the applied magnetic field. The in-plane magnetization loop derived from the series of LMOKE images in applied field is shown in Fig. 2(d).

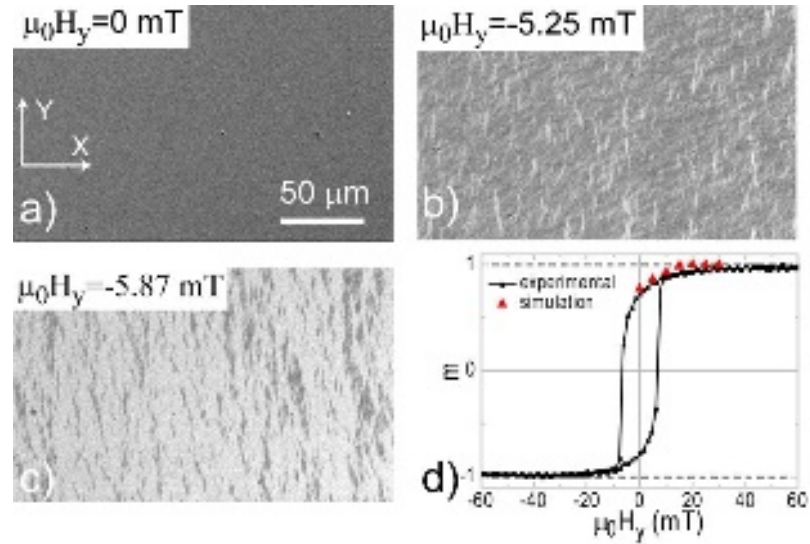

Fig. 2. (a)-(c) Series of LMOKE images of magnetic domains acquired during magnetization reversal with the in-plane magnetic field applied along the y-axis, the values of the acquisition fields are shown in the images. The darker color represents regions where the magnetization is aligned along the +y-axis, whereas the lighter color corresponds to magnetization alignment along the -y-axis. (a) Zero-field image exhibiting magnetization oriented along the +y-axis. (b) and (c) images taken slightly below and above coercive field, revealing the nucleation centers and their evolution of magnetic domains from the alignment along the +y to -y-axis. (d) Magnetization reversal loop derived from the series of LMOKE images in applied in-plane field. The data obtained from MuMax simulation are shown with red closed triangles.

Dynamical studies were performed using BLS spectroscopy in Damon-Eshbach geometry, i.e. for SWs propagating perpendicularly to the applied in-plane magnetic field. Exemplary BLS spectra (black lines) measured at magnetic field $\mu_0 H_y = -12.6$ mT (slightly smaller than coercivity field) is shown in Fig. 3(a). Stokes (negative frequency shift) and anti-Stokes (positive frequency shift) lines have been fitted by the Lorentz function to define their peak frequencies $f_S$ and $f_{aS}$. IDMI in system breaks the time-reversal symmetry of SW propagation, i.e., SWs propagating at the same frequency in the opposite directions have different wavelength and, therefore, wave number $k$ [22, 23]. This effect can be observed as a difference between $f_s$ and $f_{aS}$ peak frequencies $\Delta f = f_S - f_{aS}$ and it determines the strength of IDMI by using the equation:

$$\Delta f = \frac{2\gamma}{\pi M_S} D_{eff} k, \tag{1}$$

where $\gamma$ is gyromagnetic ratio and $M_S$ is the saturation magnetization.

In Fig. 3(a) mirror curve of Lorentz fitting (red curve) was done to calculate the $\Delta f = -1.9$ GHz. Performing the similar measurement at in-plane saturated state ($\mu_0 H_y = 300$ mT) for fixed $k=11.8$ μm$^{-1}$ and using equation (1) give us $D_{eff} = +1.74 \pm 0.12$ mJ/m$^2$ ($\gamma=176$ GHz/T and $M_S = 1210$ kA/m [24]).

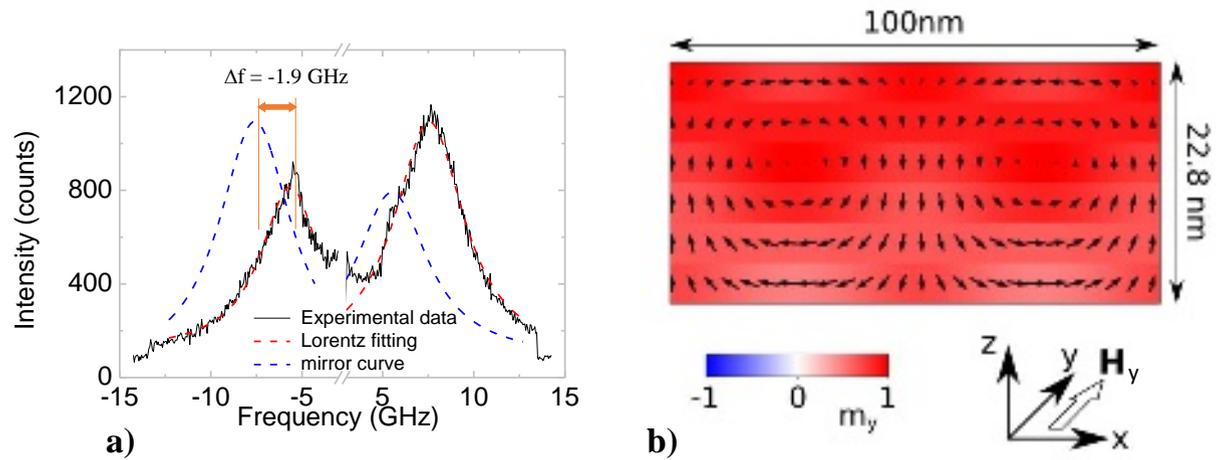

Fig. 3. (a) BLS spectrum measured for $k$=11.8 µm$^{-1}$ (black line) with Lorenz fittings (red dashed line), and its mirror curve (blue dashed line), in-plane field $\mu_0H_y$=-12.6 mT. The asymmetry between Stokes and anti-Stokes peaks $\Delta f$=-1.9 GHz is shown; (b) The simulated magnetization distribution for a multilayer consisting of six constituent magnetic films. The arrows indicate the directions of $m_x$ and $m_z$ components of magnetization, whereas the color corresponds to the $m_y$ component of magnetization.

To discuss the experimental results, micromagnetic simulations of static magnetization distribution were performed using MuMax software [24, 25]. The system consisted of 100×1×6 cells of sizes of 1×10×3.8 nm$^3$. Such an $y$-elongated shape, with periodic boundary conditions along the $x$- and $y$-axis, allow to consider the sample as infinite $xy$-plane. The sample length (100 nm) was selected to impose the experimental value of domain size period. Each magnetic film together with adjacent nonmagnetic spacers was treated as single film with total thickness of (1.8+1+1=3.8) nm, with effective magnetic parameters, as proposed by Woo at al. [25], scaled by a factor $e$=1.8/3.8 (fraction of magnetic film within the effective film). Thus the scaled experimental values of $M_S$=$e$×1.2 MA/m, $D_{eff}$=$e$×1.74 mJ/m$^2$ were used. The value of interlayer exchange scaling factor $S$ [24] was tested as a free parameter, and the value of $S$=0.015 was found, yielding the best agreement of <$m_y$> with the experimental remanence.

The cross-section view of the calculated magnetization distribution at remanence ($H$=0) is shown Fig. 3(b). The directions of magnetization are represented by the arrows and color bar defining the $m_x$, $m_z$ and $m_y$ magnetization components, respectively. In the remanent state $m_y$ component has positive value (visible as red color) resulting in high value of in-plane magnetization observed in the experiment. Analyzing the $m_x$ and $m_z$ components, one can find the domain walls resembling vortices with alternating clockwise and anticlockwise magnetization rotation around the region saturated along the $y$-direction for subsequent domain walls. These vortices separate domains with magnetization tilted towards up (+$z$) and down (-$z$) directions. The positive sign of IDMI is responsible for the shift of the magnetization core towards the upper multilayer boundary [26]. The neighbor regions with opposite out-of-plane magnetization $m_z$>0 and $m_z$<0 make the domain structure visible in MFM. On the other hand, the in-plane driven magnetization switching observed in LMOKE and VSM loops correspond to the "cores" switching between the +$y$ and -$y$ directions. With all the cores oriented along the +y direction, the shown distribution obviously represents the remanence state after the saturation by a high enough magnetic field applied along the +y direction. The simulated values of in-plane magnetization as function of applied field starting from remanence are shown in Fig. 2(d). Selected set of parameters in simulation provide good agreement with experimental data obtained from LMOKE measurements.

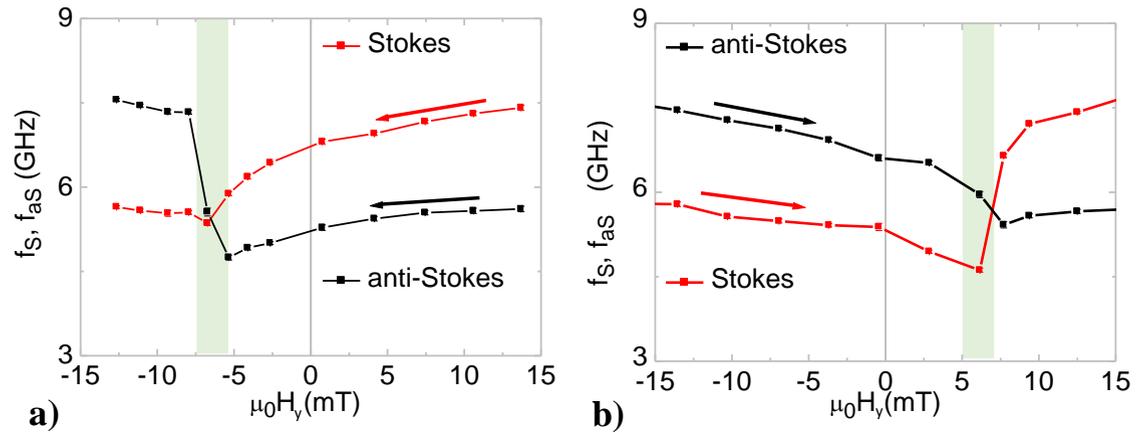

Fig. 4. (a) The dependences of the Stokes (red squares) and anti-Stokes (black squares) frequency shifts measured by BLS spectroscopy on in-plane magnetic field decreasing from +15 mT down to -15 mT. (b) the same dependence as in (a) but the magnetic field is increased from -15 mT to +15mT. The light green areas indicate applied field value when frequencies $f_S$ and $f_{aS}$ are exchanged. The arrows in the middle of the figures symbolize the direction of the field changes. The lines are guides to eyes.

BLS spectra were also studied as a function of the value of the in-plane magnetic field $H_y$. Fig. 4(a) and (b) show the Stokes $f_S$ (red squares) as well as anti-Stokes $f_{aS}$ (black circles) frequencies measured for the fixed wave number $k=11.8$ μm$^{-1}$, when $H_y$ was sweeping from positive to negative and negative to positive field directions, respectively. As the magnetic field decreases from $\mu_0H_y =15$ mT (see Fig. 4(a)) one can distinguish three field ranges: (i) $H_y > -H_C$, when $f_S > f_{aS}$; (ii) $H_y \approx -H_C$ with strong increase of $f_{aS}$ (see the green shades); (iii) $H_y < -H_C$, where $f_S < f_{aS}$. Notably, in the first region (i), the frequency shifts $f_S$ and $f_{aS}$ decrease to achieve the minimum of $f_{aS}$ and $f_S$ for the field of value $\mu_0H_y \approx$ -6.5 mT. This value corresponds well to the coercivity field $-H_C$ observed from the LMOKE magnetization curve (see Fig. 2).

As the magnetic field is swept in the opposite direction, i.e., $\mu_0H_y$ from -15 mT up to 15 mT (see Fig. 4(b)), similar three field ranges can also be distinguished. However, in this case, the switching of $f_S$ and $f_{aS}$ frequency shifts occur at the field of value $H_y\approx+H_C$. Therefore, $f_{aS} > f_S$ for fields smaller than $+H_C$, and $f_{aS} < f_S$ for fields $H_y> +H_C$.

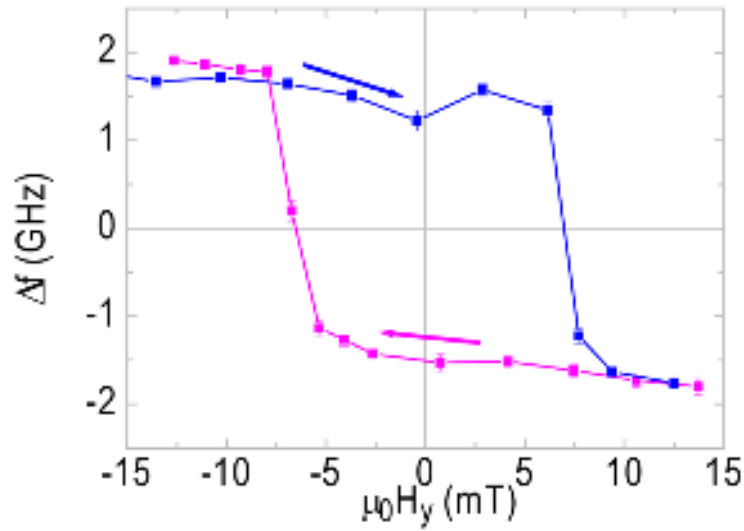

Fig. 5. Dependence of $\Delta f$ on the magnitude of in-plane applied magnetic field $H_y$. The blue squares correspond to the case with increasing value of the magnetic field $\mu_0 H_y$ from -15 mT up to 15 mT, whereas the magenta squares correspond to the case with decreasing values of the magnetic field $\mu_0 H_y$ from 15 mT down to -15 mT. The color of the arrows (blue and magenta) indicates the direction of the field changes for points of corresponding colors. The lines are guides to eyes.

The dependence $\Delta f$ ($H_y$) is shown in Fig. 5 reveals hysteretic behavior with switching at $H_C$ (similarly as switching of $<m_y>$ in Fig. 2(d)). This means that for $-H_C < H_y < H_C$ fields, depending on the history of magnetization changes, the frequency $f_{aS}$ is greater or smaller than the frequency $f_{aS}$. It is closely correlated with how the static configuration behaves with the magnetic field changes. Namely, when the sign of the overall space-averaged $m_y$ component of the magnetization ($<m_y>$) changes (in the $+H_C$ field), the sign of $\Delta f$ changes, as well. It takes negative sign ($\Delta f < 0$) for $<m_y>$ greater than 0 and positive ($\Delta f > 0$) for $<m_y> < 0$. Therefore, this switching property (i.e., different frequency of counterpropagating SWs at the same wavelengths depending on magnetization history) is a clear consequence of the IDMI presence combined together with the coercivity of the in-plane magnetization.

**Conclusions:**

In summary, the hybridization of the in-plane and out-of-plane magnetization state was studied using LMOKE microscopy and MFM for (Ir/Co/Pt)$_6$ multilayer structure, respectively. The micromagnetic simulation well supports the proposed model of these two coexisting orthogonal magnetization distributions. Exploiting BLS spectroscopy the SW propagation and their asymmetry corresponding to the wave vector orientation was investigated. Subsequently, the strength and sign of IDMI ($D_{eff} = +1.74 \pm 0.12$ mJ/m$^2$) was calculated, confirming a noticeably larger value than in the case of single Ir/Co/Pt and Pt/Co/Ir trilayers. Due to the switching of the in-plane magnetic domains switching of BLS frequencies ($f_S$ and $f_{aS}$) occurs at non-zero magnetic field (switching field) value, and it leads to a hysteresis-like behavior of frequency shift difference $\Delta f$ ($=f_S - f_{aS}$) as a function of the applied magnetic field.

This kind of system can be applicable as binary operator where at switching field two peaks Stokes and anti-Stokes interchanges it's frequencies ($f_S$ and $f_{aS}$) and intensities, and it is well guided by the magnetic history i.e. changes of the direction of in-plane applied magnetic field.

**Acknowledgement:** This work is supported by Polish National Science Center projects: DEC-2016/23/G/ST3/04196 Beethoven and UMO-2018/28/C/ST5/00308 SONATINA. And also partly supported by IEEE Magnetics Society student project 2020. PG acknowledges support by the National Science Centre of Poland, project no. 2019/35/D/ST3/03729. Micromagnetic simulations were performed in the Computational Centre of University of Bialystok.

**Data availability:** The data that support the findings of this study are available from the corresponding author upon reasonable request.

**Competing financial Interests:** The authors declare no competing financial interests.

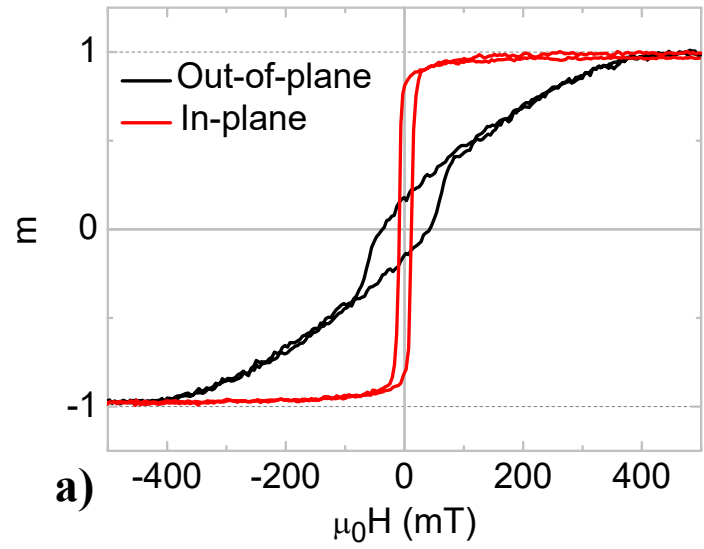 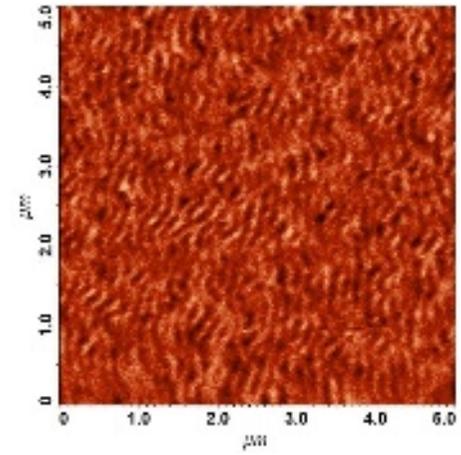

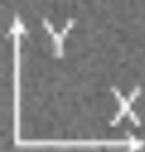
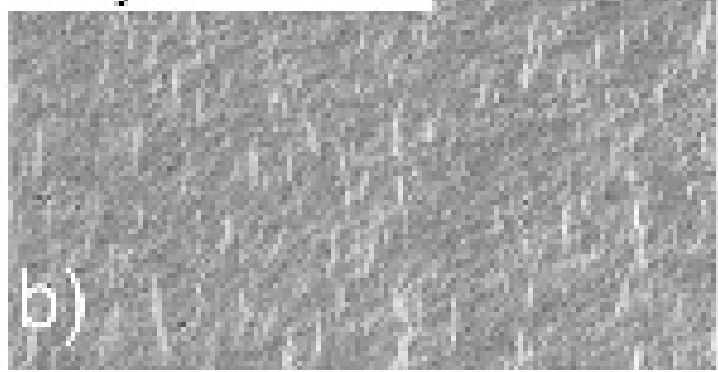
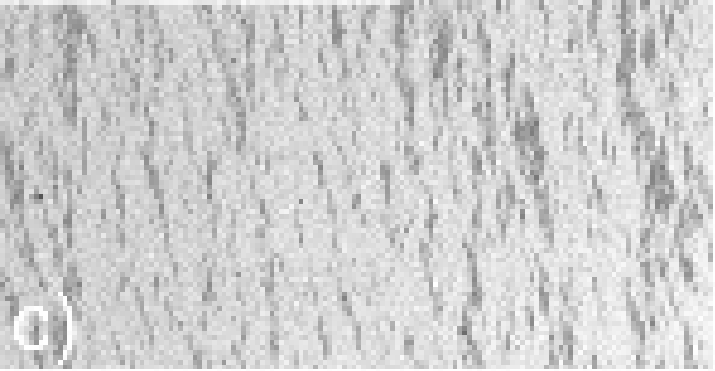
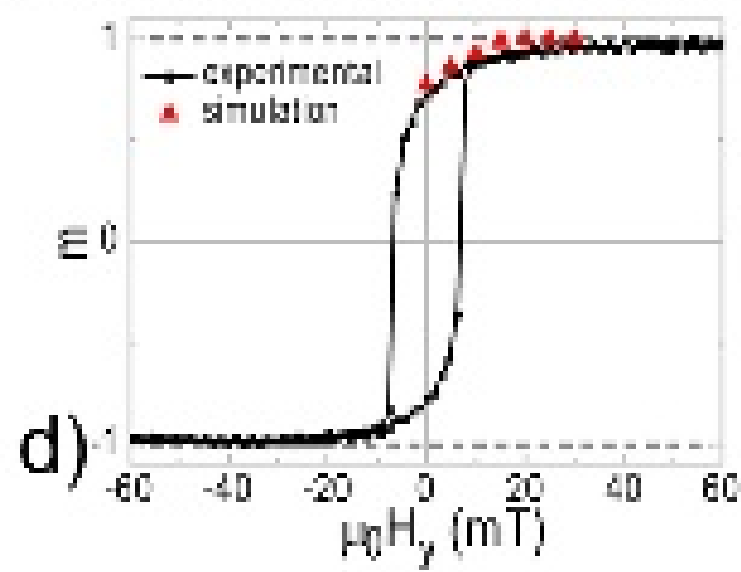

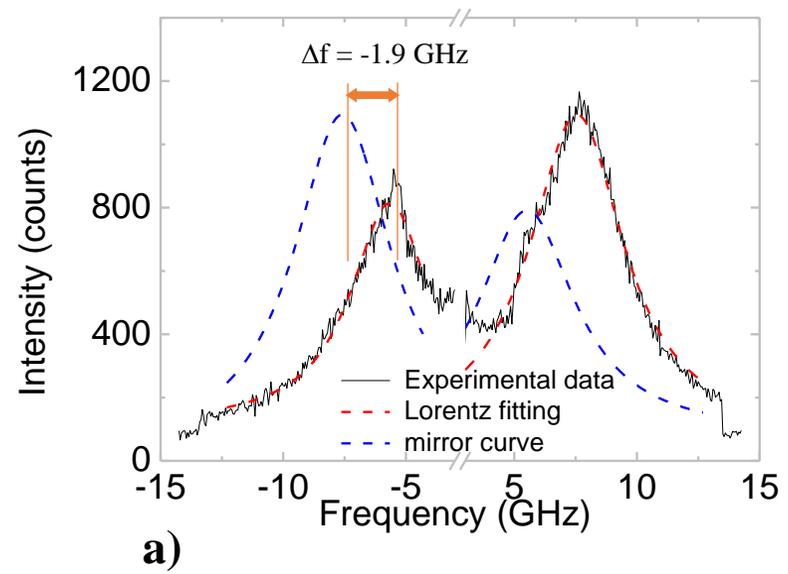 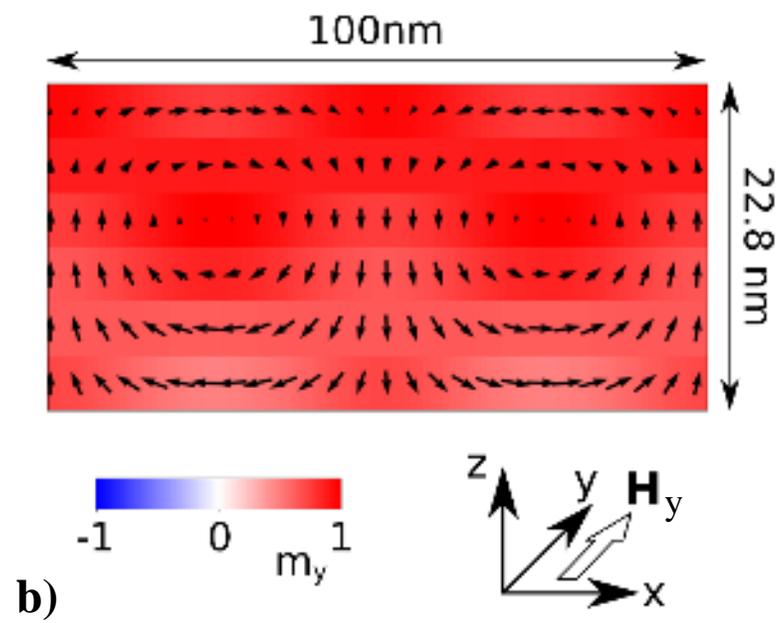

a) b)

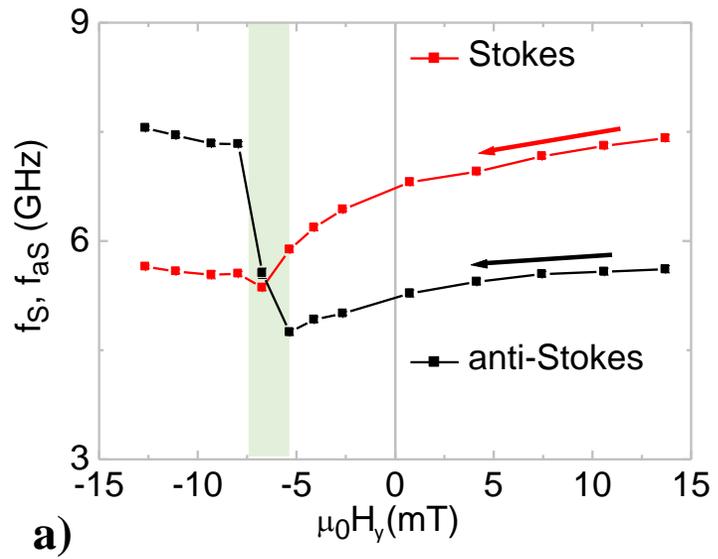 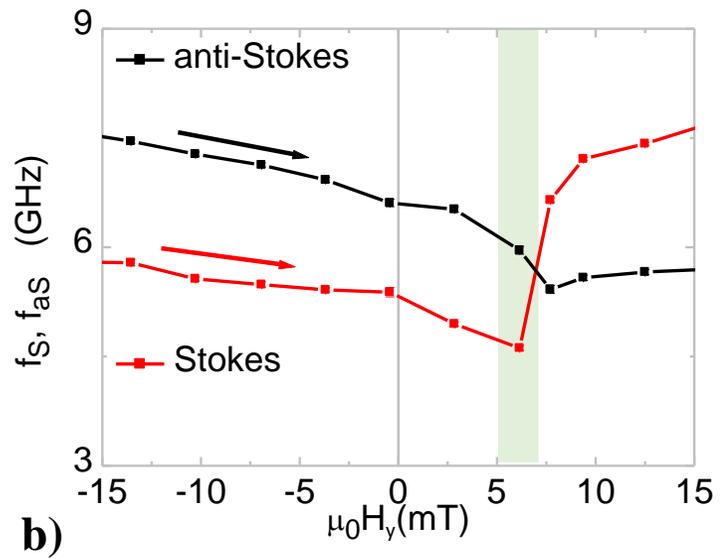

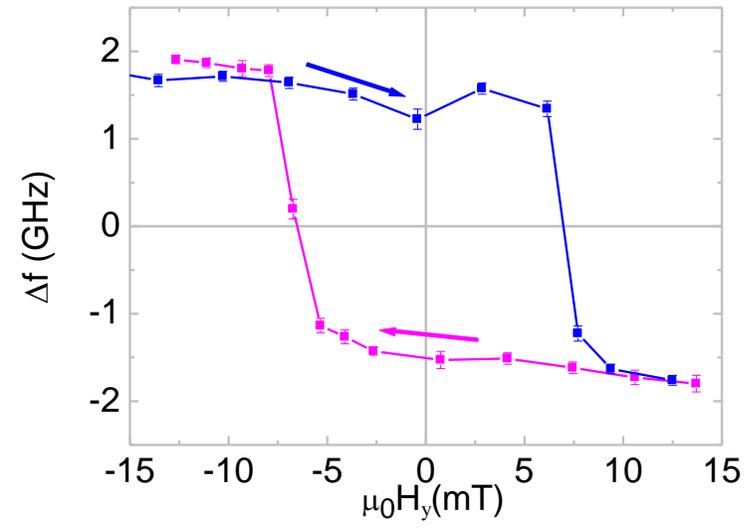